\documentclass[epj]{svjour}
\pdfoutput=1
\usepackage{amsmath}
\usepackage{amssymb}
\usepackage{graphicx}
\usepackage{epstopdf}

\begin{document}

\newcommand{\bmath}{\begin{displaymath}}
\newcommand{\emath}{\end{displaymath}}

\newcommand{\be}{\begin{equation}}
\newcommand{\ee}{\end{equation}}
\newcommand{\bea}{\begin{eqnarray}}
\newcommand{\eea}{\end{eqnarray}}
\newcommand{\non}{\nonumber\\}
\newcommand{\bmultl}{\begin{multline}}
\newcommand{\emultl}{\end{multline}}

\newcommand{\bsubeq}{\begin{subequations}}
\newcommand{\esubeq}{\end{subequations}}
\newcommand{\bitemize}{\begin{itemize}}
\newcommand{\eitemize}{\end{itemize}}
\newcommand{\ket}[1]{\left|{#1}\right\rangle}
\newcommand{\bra}[1]{\left\langle{#1}\right|}
\newcommand{\abs}[1]{\left|{#1}\right|}
\newcommand{\re}{\mathrm{Re}}
\newcommand{\im}{\mathrm{Im}}
\newcommand{\bmx}{\begin{bmatrix}}
\newcommand{\emx}{\end{bmatrix}}
\newcommand{\bsmx}{\begin{smallmatrix}}
\newcommand{\esmx}{\end{smallmatrix}}

\newcommand{\bquote}{\quotedblbase{}}
\newcommand{\equote}{\textquotedblright{ }}

\title{Self-organization of a Bose-Einstein condensate in an optical cavity}
\author{D. Nagy, G. Szirmai \and P. Domokos
}                     
%
%
\institute{Research Institute for Solid State Physics and Optics, H-1525 Budapest P.O. Box 49, Hungary}
\date{Received: date / Revised version: date}
%
\abstract{
The spatial self-organization of a Bose-Einstein condensate (BEC) in a high-finesse linear optical cavity is discussed. The condensate atoms are laser-driven from the side and scatter photons into the cavity. Above a critical pump intensity the homogeneous condensate evolves into a stable pattern bound by the cavity field. The transition point is determined analytically from a mean-field theory. We calculate the lowest lying Bogoliubov excitations of the coupled BEC-cavity system and the quantum depletion due to the atom-field coupling.
\PACS{
      {03.75.Kk}{Dynamic properties of condensates; collective and hydrodynamic excitations, superfluid flow}   \and
      {37.10.Vz}{Mechanical effects of light on atoms, molecules, and ions}
     } 
} 
\maketitle
\section{Introduction}

Atoms in a high-Q cavity, even in the dilute gas limit, form a unique many-body system due to the long-range interaction mediated by the commonly coupled cavity field mode. We have recently discovered that a gas of laser-driven cold atoms scattering light into a cavity produces a phase transition between the homogeneous spatial distribution and a regular periodic pattern \cite{domokos02b}. The ordered phase is, in fact, a `giant'  cubic crystal with a lattice constant of the optical wavelength, which is bound by the radiation field scattered by the atoms into the cavity.  This self-organization has been observed in experiment \cite{black03}, and has become a workhorse to study many-body phenomena in the strong coupling regime \cite{maschler07,larson07a,larson07b}. The realization of collective effects with material exhibiting wavelike behaviour is within reach since the first strong coupling experiments in the ultracold atom temperature regime have already been reported \cite{ottl05,brennecke07,colombe07,slama:063620}. A closely related, many-atom system in cavity is the correlated atomic recoil laser (CARL), demonstrated recently \cite{slama:053603}, which has also been studied in the ultracold gas limit \cite{piovella01}. CARL gain with BEC in free space has been observed \cite{fallani:033612}. 

In this paper we study the self-organization phe\-nom\-e\-non in the case of \emph{coupled matter and radiation fields}. As a first step in understanding the complex dynamics of this system, we neglect quantum statistical effects \cite{mekhov07a,mekhov07b} and resort to a mean field approximation.  We adopt the model developed by P. Horak et al.\  \cite{horak01b} for a different geometry.  In their scheme, the coherent laser field was injected directly into the cavity. Then the transition between homogeneous and periodically modulated densities is smooth as a function of the external pump power. By contrast, we consider the case of illuminating directly the atoms, which yields a phase transition, i.e., an abrupt change of the stationary state of the system at a well defined threshold pump power.  Note the delicate difference between stationary state and  ground state of the system. The condensate does not follow a coherent Hamiltonian dynamics because of the coupling to the decaying cavity field. The total system is dissipative owing to the irreversible photon loss from the cavity.  Thus we have to consider the steady-state of a driven open system far from equilibrium and perturbations around this steady-state.

The paper is organized as follows. In Sec.~\ref{sec:model} we present the system and the mean field model based on coupled Gross-Pitaevskii and cavity field equations. Next, in Sec.~\ref{sec:groundst} we calculate the steady-state and show that it undergoes a symmetry-breaking transition at a critical pump power. The pumping threshold for self-organization will also be discussed in terms of collective excitations in Sec.~\ref{sec:collexc}. The spectrum of low lying excitations is presented in the full range of the pumping strength in this section. The strong collective coupling regime of cavity QED is marked by a different type of transition where the ordered phase permits defect atoms in the self-organized lattice. This effect is presented in Sec.~\ref{sec:defects}. Section~\ref{sec:depletion} is devoted to studying quantum depletion which is a crucial issue concerning the stability of the condensate in a cavity and also with respect to the validity of the model.  Finally, we conclude in Sec.~\ref{sec:conclusion}.

\section{Model}
\label{sec:model}
We consider a pure Bose-Einstein condensate (BEC) interacting with a single-mode of a high-Q optical cavity. The condensate atoms are coherently driven from the side by a laser field with frequency $\omega$, directed perpendicularly to the cavity axis. The laser is detuned far from the atomic transition $\omega_A$, that is,  $|\Delta_A| \gg \gamma$, where $2\gamma$ is the full atomic linewidth at half maximum and the atom-pump detuning is $\Delta_A=\omega - \omega_A$. This condition ensures that the electronic excitation is extremely low in the condensate atoms, hence the spontaneous photon emission is suppressed.  At the same time, the laser field is nearly resonant with the cavity mode frequency $\omega_C$, i.e.\ $|\Delta_C| \approx \kappa$, where $\kappa$ is the cavity mode linewidth and the cavity-pump detuning is $\Delta_C= \omega - \omega_C$. The scattering of laser photons into the cavity is thus a qusi-resonant process and is significantly enhanced by the strong dipole coupling between the atoms and the mode due to the small  volume of the cavity. This coupling strength is characterized by the single-photon Rabi frequency $g$, which is in the range of $\kappa$. Therefore, although the condensate is hardly excited, it can efficiently scatter photons into the cavity.

For the sake of simplicity, we describe the dynamics in one dimension $x$ along the cavity axis. The cavity mode function is then simply  $\cos\,kx$. This model can apply {\it e. g.} to a  cigar shaped BEC tightly confined in the transverse directions by a strong dipole or magnetic trap, so that the transverse size of the condensate $w$ is smaller than the waist of the cavity field. The pump laser is assumed to be homogeneous along the cavity axis therefore it is described by a constant Rabi frequency $\Omega$. 

The condensate contains a number of $N$ atoms assumed to have the same wave function $\ket{\psi(t)}$.  The cavity field is assumed to be in a coherent state described by the complex amplitude $\alpha$.
These approximations imply that the quantum state of the system is factorized: entanglement between the condensate and the cavity field \cite{maschler07,vukics07b} is neglected, which can be done for large enough cavity photon number $|\alpha|^2$.  

The cavity field is subject to the strong refractive index effect of the optically dense condensate. At the same time, the evolution of the condensate wave function is described by a Gross-Pitaevskii-type equation, including the mechanical effect of the radiation field in the cavity. The system of coupled mean-field equations is
\begin{subequations}
\label{eq:motion}
\be 
i\frac{\partial}{\partial{}t}\alpha = [-\Delta_C +
N \langle{}U(x)\rangle - i\kappa]\,\alpha +
N\langle\eta_t(x)\rangle, 
\label{eq:fieldmotion}
\ee
\bea
i\frac{\partial}{\partial{}t}\psi(x,t) &=
\Big\{\frac{p^2}{2\hbar{}m} + |\alpha(t)|^2U(x) +
  2\re\{\alpha(t)\}\eta_t(x) \non
 {}& + Ng_c|\psi(x,t)|^2\Big\}\psi(x,t).
\label{eq:psimotion}
\eea
\end{subequations}
These equations are closely related to the semiclassical model for pointlike atoms \cite{domokos02b,domokos03,asboth05}, and rely on the same approximations as the  ones used in the cavity-pumping geometry \cite{horak01b}. Alternatively, this is the single-mode field case of the model of Ref.\ \cite{leonhardt99} which relies on a canonical Lagrangian approach to the action-backaction in the coupled atom-light system in the large detuning limit.  

Let us discuss the physical meaning of the terms coupling the photon
and the matter fields. Each atom shifts the cavity resonance frequency
in a spatially dependent manner by $U(x)=U_0 \cos^2{kx}$.  The maximum
shift is $U_0 = g^2/\Delta_A$, obtained at the antinodes of the mode
function.  In the mean-field approximation, the shift has to be
spatially averaged over the single-atom wave function, giving the
frequency shift per atom $\langle{}U(x)\rangle= \, U_0 \bra{\psi}
\cos^2{kx}\ket{\psi}$.  It is worth noting that, even if the one-atom
light shift $U_0$ is small compared to the cavity decay $\kappa$, one
can achieve the interesting strong collective coupling regime of
cavity QED, $N|U_0| > \kappa$, with a trapped BEC.  Feeding the cavity
by laser scattering on the atoms appears as an effective pump with
strength $\eta = \Omega g/\Delta_A$. This process has a spatial
dependence inherited from the mode function, $\eta_t(x)= \eta \cos kx$, and this term also has to be averaged  over the condensate wave function.

The well known Gross-Pitaevskii equation (GPE) in one dimension in Eq.~(\ref{eq:psimotion}) describes the motion of a scalar condensate induced by the optical forces in the cavity field. The back action of the light shift is the term proportional to \mbox{$U(x) = U_0\cos^2{kx}$}
and the cavity photon number $|\alpha|^2$. This term is periodic with half of the wavelength $\lambda$ and, in free space, is referred to as an ``optical lattice''. The back action of the coherent scattering of photons between  the transverse pump and the cavity mode is the term proportional to 
$\eta_t(x) = \eta\cos\,kx$ and has a periodicity of $\lambda$.
The last term of the GPE accounts for the s-wave collision of the 
atoms, its strength is related to the s-wave scattering length $a$ 
by \mbox{$g_c = 4\pi\hbar{}a/(mw^2)$}, and depends on the
transverse size of the condensate $w$.

In this paper we will consider the case $U_0 < 0$, {\it i. e.} large
red detuning, where the atoms behave as high field seekers. 
Consequently, for nonzero cavity field the condensate atoms tend to localize 
around the field antinodes, thereby maximizing their coupling to the light field.

\subsection{Scaling with the atom number}

It can be checked that Eqs.\ (\ref{eq:motion}) with the field amplitude replaced by $\alpha/\sqrt{N}$ are invariant under the scaling of the parameters such that $N U_0$, $N g_c$, and $\sqrt{N} \eta$ is kept constant. That is, in the mean field model, the atom number can be incorporated in the system parameters and the field amplitude variable (with the proposed rescaling, the absolute square of this latter gives the photon number per atom). Therefore, in the rest of this paper the system parameters will occur only in the form of the above combinations. However, we refrain from introducing a new notation for the scaled parameters in order to signify their relation to experimental parameters.

Further simplification for the numerical method can be obtained by making use of the periodicity of the optical potential.  We can consider periodic wave functions $\psi(x)$, and solve Eqs.~(\ref{eq:statsol}) in the interval $[0,\lambda]$ using periodic
boundary condition. In this way, we discard the dynamics within a ``Bloch band''. We are interested in effects due to  the condensate-cavity field interaction which does not couple states of different quasi momenta. Thence it is independent of the length of the interval and we choose the shortest one to reduce the computational effort. By contrast, the collisional interaction depends on the actual atom density, hence, for a fixed atom number $N$, the density is artificially enhanced by folding the space into a single $\lambda$ interval. Therefore, the collision parameter has to be modified such that $N g_c/\lambda$ correspond to the collisions at the actual (experimental) atom density in the cavity.

\section{Steady state}
\label{sec:groundst}

In this section, we intend to study the steady state of the compound condensate-cavity system. It is described by the field amplitude $\alpha_0$, and the condensate wave function $\psi(x, t) = \psi_0(x)e^{-i\mu{}t}$, where $\hbar\mu$ is the chemical potential. Following from Eq.\ (\ref{eq:motion}), they obey the system of equations
\bsubeq
\label{eq:statsol}
\be
\label{eq:alfa0}
\alpha_0 = \frac{N\bra{\psi_0}\eta_t(x)\ket{\psi_0}}{
  \Delta_C - N\bra{\psi_0}U(x)\ket{\psi_0} + i\kappa} \; ,
\ee
\begin{multline}
\label{eq:psi0}
\bigg\{\frac{p^2}{2\hbar{}m} + \;|\alpha_0|^2U(x) 
+ 2\re\{\alpha_0\}\eta_t(x) \\ + Ng_c|\psi_0(x)|^2\bigg\} \psi_0(x) = \mu\psi_0(x) \; .
\end{multline}
\esubeq
The solution for $\psi_0(x)$, $\mu$ and $\alpha_0$ can be determined
in a self-consistent manner. We use a variant of the imaginary time
propagation method.  Our method consists of propagating the condensate
wave function in imaginary time $\tau = it$ according to
Eq.~(\ref{eq:psimotion}) complemented by the adiabatic elimination of
the cavity field dynamics. The field amplitude is expressed in terms
of the instantaneous wave function $\psi(x)$, similarly to the form of Eq.~(\ref{eq:alfa0}), and inserted into Eq.~(\ref{eq:psimotion}) in each time step. Starting from a guessed initial
function, the wave function $\psi(x)$ decays with time 
$\tau$. It is just the solution $\psi_0$ that undergoes the
slowest decay, because it has the smallest energy $\hbar\mu$. Components
from the excited states with higher energy decay faster. Therefore, by
renormalizing $\psi(x)$ in each step to have a norm 1, the iterations converge to the solution $\psi_0(x)$.
This solution is a self-consistent ground state of the condensate in the optical potential corresponding to $\alpha_0$.

\subsection{Brief summary on cavity-induced self-organization}
\label{subsec:selforg}
The mean-field model of the condensate is closely related to  the mean-field model of a thermal gas of atoms presented in \cite{asboth05,nagy06}. Therefore we can expect that the phase transition appears as an abrupt change of the self-con\-sist\-ent steady state at a critical pumping strength $\eta$. The physical process is analogous, however, at zero temperature the kinetic energy and the atom-atom collisions are apt to stabilize the phase with larger symmetry. In the following, we briefly summarize the basic elements of self-organization.

In order to gain some physical insight, it is appropriate to adiabatically eliminate the cavity field. The resulting  self-consistent potential in the GPE (\ref{eq:psimotion}) is
\be 
\label{eq:adpot}
V(x) = U_1\cos{kx} + U_2\cos^2{kx} \; , 
\ee 
which is the sum of a $\lambda$ and a $\lambda/2$ periodic potential. The coefficients are 
\bsubeq
\be 
\label{eq:U1}
U_1 = 2 \langle\cos{kx}\rangle\,  N I_0\, [\Delta_C - 
NU_0\langle\cos^2{kx}\rangle] \; ,  
\ee 
\be
\label{eq:U2}
U_2 = \langle\cos{kx}\rangle^2\, N^2 I_0 \,U_0\; .
\ee 
\esubeq 
The nonlinearity of the system is introduced by the dependence of the coefficients on the wave function via the specific mean values
\be 
\Theta = \bra{\psi_0}\cos{kx}\ket{\psi_0}, 
\ee 
which can be considered as an \emph{order parameter}, and
\be 
\mathcal{B} = \bra{\psi_0}\cos^2{kx}\ket{\psi_0}, 
\ee 
which is called the  \emph{bunching parameter}. The order parameter describes the 
$\lambda$-periodic spatial order of $\psi_0$: $\Theta = 0$ for a 
uniform distribution, and $\Theta = \pm 1$ for the atoms being
localized around the `even' ($kx = 2n\pi$), or `odd' field antinodes ($kx = (2n+1)\pi$), respectively.  
The bunching parameter reflects the degree of localization of the 
atoms into the potential wells of the optical lattice. Finally, 
$I_0$, scaling the depth of the potential, represents the maximum number of photons an atom can scatter 
into the cavity 
\be 
\label{eq:I0}
I_0 = \frac{|\eta|^2}{[\Delta_C - NU_0\mathcal{B}]^2 + \kappa^2}.
\ee

It can be easily seen that $\psi_0(x) \equiv 1$ and $\alpha_0 = 0$ is a trivial solution of Eqs.~(\ref{eq:statsol}). 
Because $\langle\cos{kx}\rangle = 0$ for uniform atomic distribution, the cavity field driving term vanishes and $\alpha_0 = 0$. There is no potential then, $U_1=U_2=0$, thus the uniform distribution $\psi_0(x) = 1$  remains a good solution. However, it may not be a stable self-consistent ground state of the condensate, because fluctuations of the wave function can be amplified for adequately chosen parameters.

Self-organization is based on the effect of the $\lambda$ periodic $U_1$ potential term which involves a positive feedback mechanism.  Let us set the detuning $\Delta_C < -N|U_0|$ so that the
sign of $U_1$ becomes the opposite of the sign of $\Theta$, {\it cf.}\ Eq.\ (\ref{eq:U1}). Consequently, if some $\lambda$-periodic fluctuation of the condensate yields $\Theta > 0$ ({\it i.e.}, more
atoms happen to be near the even sites than near the odd ones), it produces
a potential $U_1 \cos{kx}$ (with $U_1 < 0$) which has minima at the even
sites, thus attracting even more atoms there. Simultaneously, the $\lambda$-periodic optical lattice of
condensate atoms fulfills the Bragg-condition for constructive interference, $\langle \eta_t(x) \rangle \neq 0$, so the atoms can scatter pump photons into the cavity. The positive feedback requires then just
that the parameters $U_1$ and $\Theta$ have opposite signs. This can be ensured  by choosing the cavity detuning
\be
\label{eq:DC}
\Delta_C = NU_0 - \kappa\; ,
\ee
which is a sufficient but not necessary condition (because the bunching parameter is less than 1,  the field could be tuned closer to resonance).

The $\lambda/2$ periodic $U_2$ potential term does not discriminate between the even and odd sites. Moreover, it is proportional to the square of the order parameter, thus it plays no role in the onset of self-organization. The consequences of this potential term will be discussed later.

The runaway solution is counteracted by the kinetic energy and the
collisions, both trying to maximally spread the atomic wave function.
There is a critical pump power above which the self-organization can
occur.

\subsection{Results from the Gross-Pitaeskii equation}
\label{subsec:results}

The relevant parameters are measured in units of the recoil frequency
$\omega_R = \hbar{}k^2/(2m)$ and in units of the recoil energy $\hbar\omega_R$,
respectively. The natural length scale is the optical wavelength $\lambda$ of
the cavity field.  We set the strength of the s-wave
interaction $N g_c =10\,\omega_R\lambda$.
The atom-cavity coupling is characterized by $NU_0 = 100\,\omega_R$,
which is on the order of the cavity decay rate, $\kappa =
200\,\omega_R$. We set the value of the cavity detuning, according to
Eq.~(\ref{eq:DC}), $\Delta_C = -300\,\omega_R$.

\begin{figure}
\begin{center}
\resizebox{0.85\columnwidth}{!}{%
\rotatebox{0}{\includegraphics{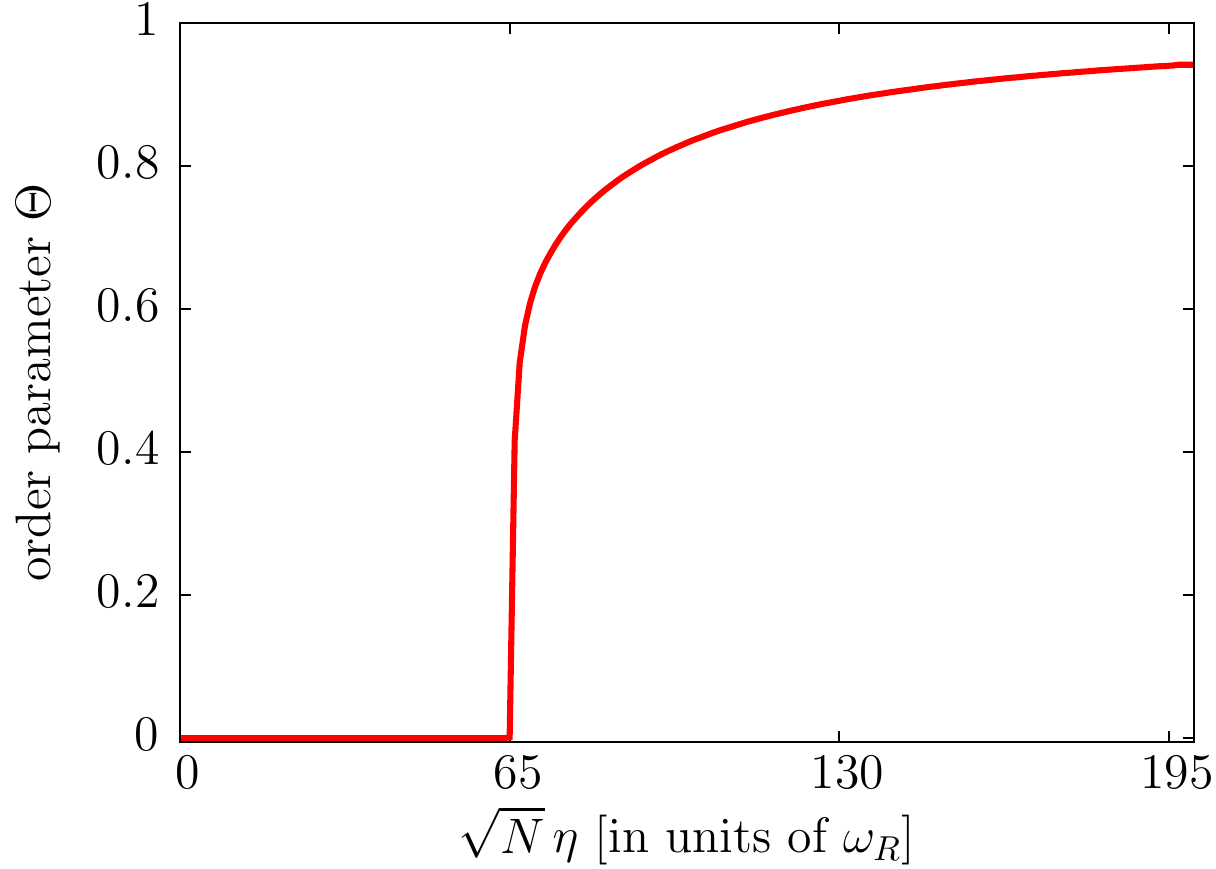}}}
\end{center}
\caption{The order parameter $\Theta$, plotted as a function of the
  transverse pump amplitude $\sqrt{N}\eta$, exhibits the
  self-or\-gan\-i\-za\-tion of the BEC into a $\lambda$-periodic optical
  lattice. Parameters: $N g_c =10 \hbar{}\omega_R\lambda$, $N U_0 =
  -100 \omega_R$, $\Delta_C = -300\omega_R$, $\kappa = 200\omega_R$.}
\label{fig:theta_eta}       
\end{figure}

The numerical solution for the self-consistent ground state of the condensate-cavity system confirms the qualitative picture presented above. In Fig.~\ref{fig:theta_eta}, the appearance of the self-organized lattice is manifested by the variation of the order parameter $\Theta$ against the pumping strength $\sqrt{N} \eta$.  

\begin{figure}
\begin{center}
\resizebox{0.90\columnwidth}{!}{%
\rotatebox{90}{\includegraphics{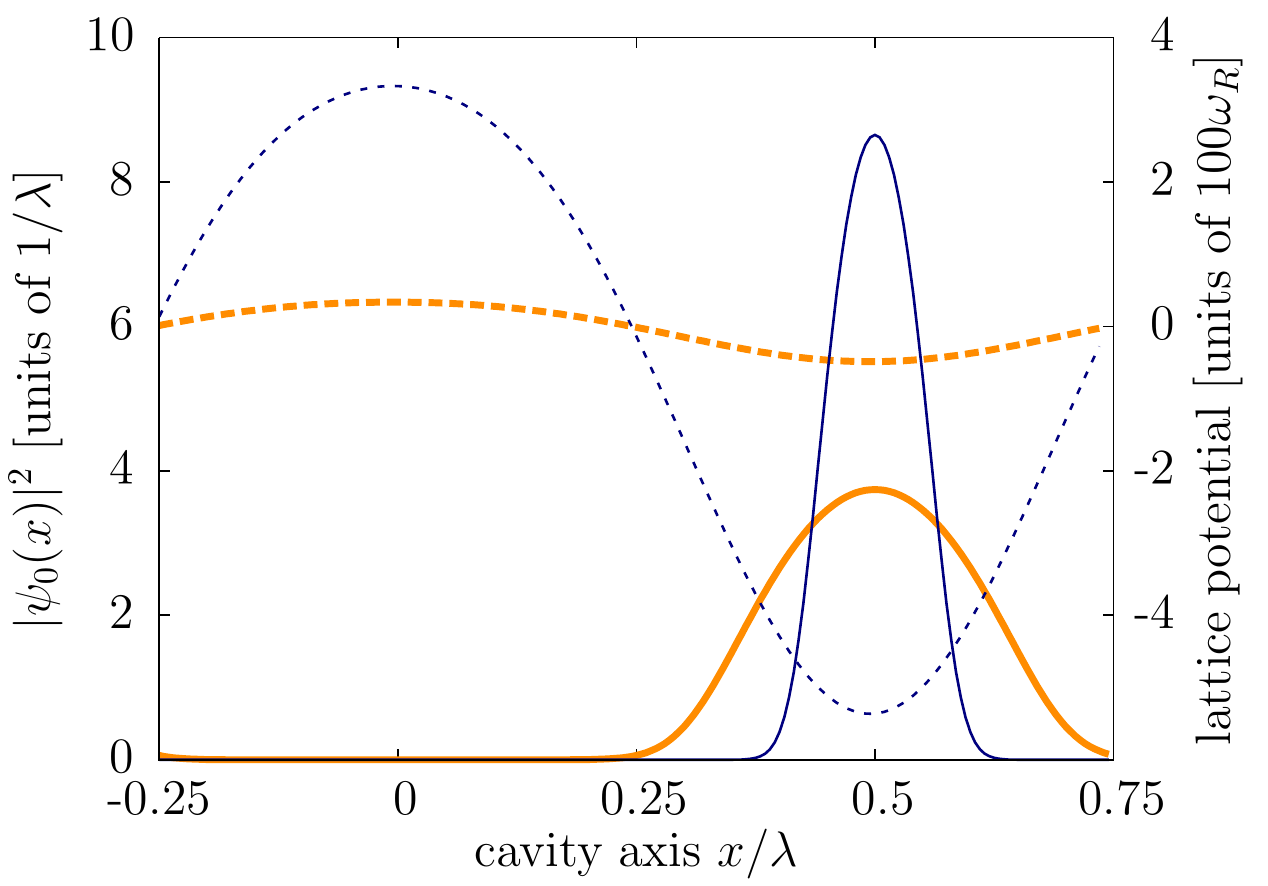}}}
\vspace{.3cm}
\end{center}
\caption{Two typical atomic position distibutions  $|\psi_0(x)|^2$ (solid lines) and the corresponding adiabatic 
optical potentials from Eq.~(\ref{eq:adpot}) (dashed lines) for pump strengths $\sqrt{N}\eta = 100\omega_R$ (thick lines) and {$\sqrt{N}\eta = 300 \omega_R$ (thin lines)}. Other parameters are the same as in 
Fig.~\ref{fig:theta_eta}.}
\label{fig:loc}       
\end{figure}

Above a critical pump amplitude ($\sqrt{N} \eta_c \approx 65.6\,\omega_{R}$), the homogeneous condensate, our initial guess for the condensate wave function, self-organizes into a lattice of period $\lambda$.  At the outset of the self-organization, a spontaneous symmetry breaking occurs: the condensate occupies either the $k x=0$ site or the $x=\pi$ site. Further increasing the pump strength $\sqrt{N}\eta$, the atoms get more and more localized around the chosen site, that is indicated by the growth of the order parameter $\Theta$.  The localization of the condensate  in the self-organized phase is showed in Fig.~\ref{fig:loc} for two specific transverse pump amplitudes $\sqrt{N}\eta = 100$ (thick lines) and $300\,\omega_R$ (thin lines).

This localization behavior can remind the reader of a superfluid-Mott (SM) phase transition (see e.g. Ref. \cite{bloch2008} for a review). An SM transition happens in a deep enough optical lattice, where the competition between kinetic energy and on-site interaction results in distinct ground-state configurations. If the on-site interaction dominates, the multiple occupation of a single site is energetically disfavored and the ground state corresponds to  localized atoms evenly distributed over the lattice points (Mott phase). Contrarily, when the kinetic energy dominates, the zero quasi momentum state (completely delocalized) is macroscopically occupied (superfluid phase). This shows similarity to the self-organization discussed above, with the difference that the competition now is between the kinetic energy and the energy of the cavity field. This difference, however, leads to some other distinctions between the two transitions. In the case of the self-organization in an cavity the optical potential builds up for $\eta>\eta_c$ and is shallow in the beginning. Therefore the macroscopic wave function, although it breaks the continuous translational invariance, is not strongly localized (at least for $\eta\approx\eta_c$). Even if the optical potential is deep ($\eta\gg\eta_c$) the compressibility of the system, $\partial N/\partial \mu$, is nonzero, which is another difference compared to the SM transition. The role of the interparticle interaction strength is also opposed in the self-organization transition, since it spreads out the macroscopic wave function and counteracts localization. It is also important to note, that even in the deeply localized limit of the self-organized phase, the number of bosons in a single site can be large (depending on the parameters of the actual realization) and the Gross-Pitaevskii approximation applies. For example, in the experiments made by the Esslinger group \cite{brennecke07}, the number of condensed atoms were approximately $N=4\cdot10^5$, while the size of the condensate were about $30\lambda$. Therefore there can be about $10^4$ atoms per lattice site in a possible experimental realization, which is a huge number compared to those found in a Mott phase. 

The critical pump amplitude $\sqrt{N}\eta_c$ can be determined
analytically by the stability analysis of the trivial solution of
Eqs.~(\ref{eq:statsol}), being $\psi_0(x) \equiv 1$, $\alpha_0 = 0$,
$\mu_0 = Ng_c$. Only the Fourier component $\cos\,kx$ of a noisy
perturbation of the wave function can produce a non-vanishing cavity field $\alpha_0$ other than zero, in accordance with the mean value $\bra{\psi}\cos\,kx\ket{\psi}$ in the numerator of Eq.~(\ref{eq:alfa0}). Therefore, starting from the wave function $\psi(x) = 1 + \epsilon \cos\,kx$, with $\epsilon \ll 1$, we carry out one iteration step of the imaginary time propagation method: 
\begin{align}
\frac{\Delta\psi}{\Delta\tau} = &-  Ng_c -  \epsilon\cos{kx} \times\nonumber\\
  & \left(\omega_R + N\eta^2\frac{2\Delta_C - NU_0} {(\Delta_C - NU_0/2)^2 + \kappa^2} + 3Ng_c\right) \; .
\end{align}
The component $\psi_0 = 1$ decays with the rate 
$\mu_0 = Ng_c$, while the decay of the perturbation $\cos\,kx$ depends 
on the pumping strength $\eta$. In order to preserve the stability of the homogeneous ground state,
the perturbation should decay faster than $\psi_0$. The
condition that the coefficient of $\epsilon\cos\,kx$ is equal to $Ng_c$ leads to the critical pump amplitude
\be
\label{eq:eta_c}
\sqrt{N}\eta_c = \sqrt{\frac{(\Delta_C - NU_0/2)^2 +
    \kappa^2}{(NU_0 - 2\Delta_C)}}\sqrt{\omega_R + 2Ng_c}\; .  
\ee
Interestingly, the critical pump amplitude given by
Eq. (\ref{eq:eta_c}) is formally analogous to the one obtained in the case of a thermal classical
gas \cite{nagy06}.  The temperature is substituted by the kinetic energy and the s-wave collision of the atoms, formally $k_BT/\hbar \leftrightarrow \omega_R + 2Ng_c$.

\section{Collective excitations}
\label{sec:collexc}
Let us now calculate the excitation spectrum of the coupled condensate-cavity system as the linear response of the self-consistent steady-state. Two limiting cases can be relatively easily understood: (i) for $\eta=0$ there is no field in the cavity and one gets back the excitation spectrum of a homogeneous Bose-gas; (ii) for $\eta \rightarrow \infty$, deeply in the self-organized phase, where the optical potential can be approximated as a parabola, one expects to obtain the excitations of a BEC in harmonic trap potential. This simplification, however, does not perfectly apply since the excitations are not only collective oscillations of the atom cloud, but they are \emph{polaritons}  involving the fluctuation of the field amplitude around its steady state. In the following, we will consider the full transition range, including the critical point,  between these limiting cases.

We need to consider the deviations from the stationary state ($\psi_0$ and $\alpha_0$): 
\bsubeq
\be
\alpha(t) = \alpha_0 + \delta\alpha(t) \; ,
\ee
\be
\psi(x,t) = e^{-i\mu{}t}[\psi_0(x) + 
\delta\psi(x,t)]\; .
\ee
\esubeq
Inserting the ansatz into Eqs.~(\ref{eq:motion}) and linearizing in 
$\delta\psi$ and $\delta\alpha$, one gets
\bsubeq
\label{eq:dmotion}
\begin{multline}
\label{eq:dalpha}
i\delta\dot\alpha = A\:\delta\alpha 
+ N\alpha_0[\bra{\psi_0}U(x)\ket{\delta\psi} + \bra{\delta\psi}U(x)\ket{\psi_0}] \\
+ N[\bra{\psi_0}\eta_t(x)\ket{\delta\psi} + \bra{\delta\psi}\eta_t(x)\ket{\psi_0}]\; ,
\end{multline}
\bea
\label{eq:dpsi}
i\delta\dot\psi &= \left\{H_0 + Ng_c|\psi_0(x)|^2\right\}\delta\psi 
+ Ng_c\psi_0^2(x)\delta\psi^{*} \non
&+ \psi_0(x)U(x)(\alpha_0\delta\alpha^{*} + \alpha_0^{*}\delta\alpha) \non
&+ \psi_0(x)\eta_t(x)(\delta\alpha + \delta\alpha^{*})\; ,
\eea
\esubeq
where
\bsubeq
\be 
A = -\Delta_C + N \bra{\psi_0}U(x)\ket{\psi_0}-i\kappa \; ,
\ee 
and
\begin{multline}
H_0 = \frac{p^2}{2\hbar{}m}+ Ng_c|\psi_0(x)|^2-\mu\\ + |\alpha_0|^2U(x) + 2\re\{\alpha_0\}  \eta_t(x)  \; .
\end{multline}
\esubeq
Because the linearized time evolution couples
$\delta\psi$ and $\delta\alpha$ to their complex conjugates, we search the solution in the form
\bsubeq
\label{eq:linearizedGPE}
\be
\delta\alpha(t)=e^{-i\omega t}\delta\alpha_++e^{i\omega^{*}t}\delta\alpha_-^{*},
\ee
\be
\delta\psi(x,t) = e^{-i\omega t}\delta\psi_+(x) + 
e^{i\omega^{*}t}\delta\psi^{*}_-(x),
\ee
\esubeq
where $\omega=\nu-i\gamma$ is a complex parameter of the oscillation 
standing for frequency $\nu$ and damping rate $\gamma$. Equations~(\ref{eq:dmotion}) have to be obeyed separately for the $e^{-i \omega t}$ and $e^{i\omega^{*}t}$ terms, which leads to the linear eigenvalue equation:
\be
\label{eq:eigenproblem}
\omega \begin{pmatrix}
        \delta\alpha_+ \\
        \delta\alpha_- \\
        \delta\psi_+(x) \\
        \delta\psi_-(x) \\
       \end{pmatrix} = \mathbf{M} 
       \begin{pmatrix}
        \delta\alpha_+ \\
        \delta\alpha_- \\
        \delta\psi_+(x) \\
        \delta\psi_-(x) \\
       \end{pmatrix},
\ee
where $\bf M$ is a non-Hermitian matrix being determined by Eqs.~(\ref{eq:dmotion}). Collective excitations of the system are the solutions of this eigenvalue problem.
It simplifies if we choose $\psi_0(x)$ real, and 
write $\bf M$ in the basis of the symmetric and antisymmetric 
combinations, 
\begin{equation}
\begin{split}
\delta\alpha_a &= \delta\alpha_+ - \delta\alpha_- \; ,\\
\delta\alpha_s &= \delta\alpha_+ + \delta\alpha_-\; , \\
\delta f(x) &= \delta\psi_+(x) + \delta\psi_-(x) \; , \\
\delta g(x) &= \delta\psi_-(x) - \delta\psi_+(x)\; .
\end{split}
\end{equation}
Transformed into this basis, the  matrix $\bf M$ is  
\be
\label{mx:M}
\!\bmx
-i\kappa & \re{A} & 2N(\re{\alpha_0}{\rm X} + {\rm Y}) & 0 \\
\re{A}  & -i\kappa& 2iN\im{\alpha_0}{\rm X} & 0 \\
 0      &    0   &      0                  & -H_0 \\
 2i\psi_0U\im\alpha_0 & -2\psi_0(U\re{\alpha_0} + \eta_t) 
 & -H_0-2Ng_c\psi_0^2 & 0
\emx\!.
\ee
The argument $x$ has been omitted from the functions $\psi_0$, $U$ and $\eta_t$ 
for brevity. The integral operators $X$ and $Y$, coupling the condensate excitations into 
the field, read
\bsubeq
 \label{mx:coupling}
\be
{\rm X}{\xi(x)} = \int{}dx\:\psi_0(x)U(x)\xi(x),
\ee
\be
{\rm Y}{\xi(x)} = \int{}dx\:\psi_0(x)\eta_t(x)\xi(x).
\ee
\esubeq
Being a non-Hermitian matrix, $\bf M$ has different right and
left eigenvectors corresponding to the same eigenvalue $\omega$. 

The matrix has a special symmetry: if $\omega$ is an
eigenvalue with the right eigenvector $$(\delta\alpha_a,
\delta\alpha_s, \delta{}f, \delta{}g)\,,$$ then $-\omega^*$ is also an
eigenvalue with the right eigenvector $$(-\delta\alpha_a^*,
\delta\alpha_s^*, \delta{}f^*, -\delta{}g^*)\,.$$ Thus the eigenvalues
come in pairs,  having the same imaginary parts but the real parts are of the opposite sign in a pair.  This grouping of the eigenvectors makes sense for eigenvalues with non-vanishing real part. 

The imaginary part describes damping which arises from the nonadiabaticity of the cavity field dynamics.  The linear perturbation calculus of the excitations which we adopted here takes into account that the field follows the changes of the BEC wave function with a delay in the order of $1/\kappa$. Depending on the specific choice of the parameters, this can yield damping or heating, which is known as cavity cooling and has been extensively studied in the past decade. Damping of BEC excitations in optical cavities, which has been first studied in Refs.\ \cite{horak01b,gardiner:051603}, is an interesting opportunity which motivates the realization and the study of the coupled cavity-condensate system.

\subsection{Spectrum analytically below threshold}

Below threshold, the condensate wave function is constant and linearization around this simple solution leads to analytical results. The stability of the uniform distribution can be evaluated from the spectrum, thus the critical point can be determined in this way, independently of the previous calculation in Sec.~\ref{subsec:results}. Moreover, we obtain a detailed description of a restricted part of the spectrum which contains polariton excitations. 

Assuming  homogeneous atomic distribution and, correspondingly, vanishing cavity field ($\psi_0\equiv 1$, $\alpha_0 = 0$), the only non-trivial coupling term in the matrix $\bf M$, see Eq.\ (\ref{mx:M}), is the $Y$ operator defined in Eq.\ (\ref{mx:coupling}b). Since $\eta_t(x)\propto \cos{k x}$, only the Fourier component $\cos\,kx$ couples to the cavity amplitude. In return, the fourth line of the matrix $\bf M$ shows that the field fluctuations excite just this $\eta_t(x)\propto\cos{kx}$ condensate perturbation. Thus, this subspace is closed below threshold. All the other condensate excitations decouple from the field and remain simply the higher Fourier components with frequencies 
\be
 \label{eq:bogoliubov}
\Omega_n = \sqrt{n^2\omega_R(n^2\omega_R + 2Ng_c)}\; , \quad (n>1)\;, 
\ee 
which is identical to the excitations of a homogeneous condensate in a box \cite{castin}.  It is sufficient to diagonalize the matrix $\bf M$ in the restricted subspace, 
\be
\label{mx:Mcos}
\bmx
-i\kappa & \delta_C & N\eta & 0 \\
\delta_C  & -i\kappa& 0 & 0 \\
 0      &    0   &      0   & -\omega_R \\
 0      & -2\eta & -\omega_R-2Ng_c & 0
\emx \,.
\ee
It has  the fourth order characteristic equation
\be
\label{eq:kar}
(\lambda^2 - \Omega_1^2)\left[(i\kappa + \lambda)^2 - \delta_C^2\right]
- 2N\eta^2\omega_R\delta_C = 0,
\ee
where $\delta_C = -\Delta_C + \frac{1}{2}NU_0$, and $\Omega_1$ is the first excitation energy in the Bogoliubov spectrum (\ref{eq:bogoliubov}) of a BEC in a box. 
One can check  that, for \mbox{$\eta = 0$}, the last term vanishes and $\lambda_{1,2} = \pm \Omega_1$ for the condensate excitation,
$\lambda_{3,4} = \pm\delta_C - i\kappa$ for the cavity mode. For non-zero pumping, $\eta\neq 0$, the condensate-like and the field-like excitations mix. When $\Omega_1^2 \ll  \kappa^2 +\delta_C^2$, which case we consider in this paper, the frequencies corresponding to the excitations of the free field and to the free condensate are well separated. Therefore, the mixing ratio is small so that the polariton modes can be attributed to dominantly condensate or field excitations.

At the onset of self-organization the uniform ground state $\psi_0$ changes, which, in the present approach, corresponds to the appearance of a zero eigenvalue in the spectrum. 
\begin{figure}
\begin{center}
\resizebox{0.9\columnwidth}{!}{%
\rotatebox{0}{\includegraphics{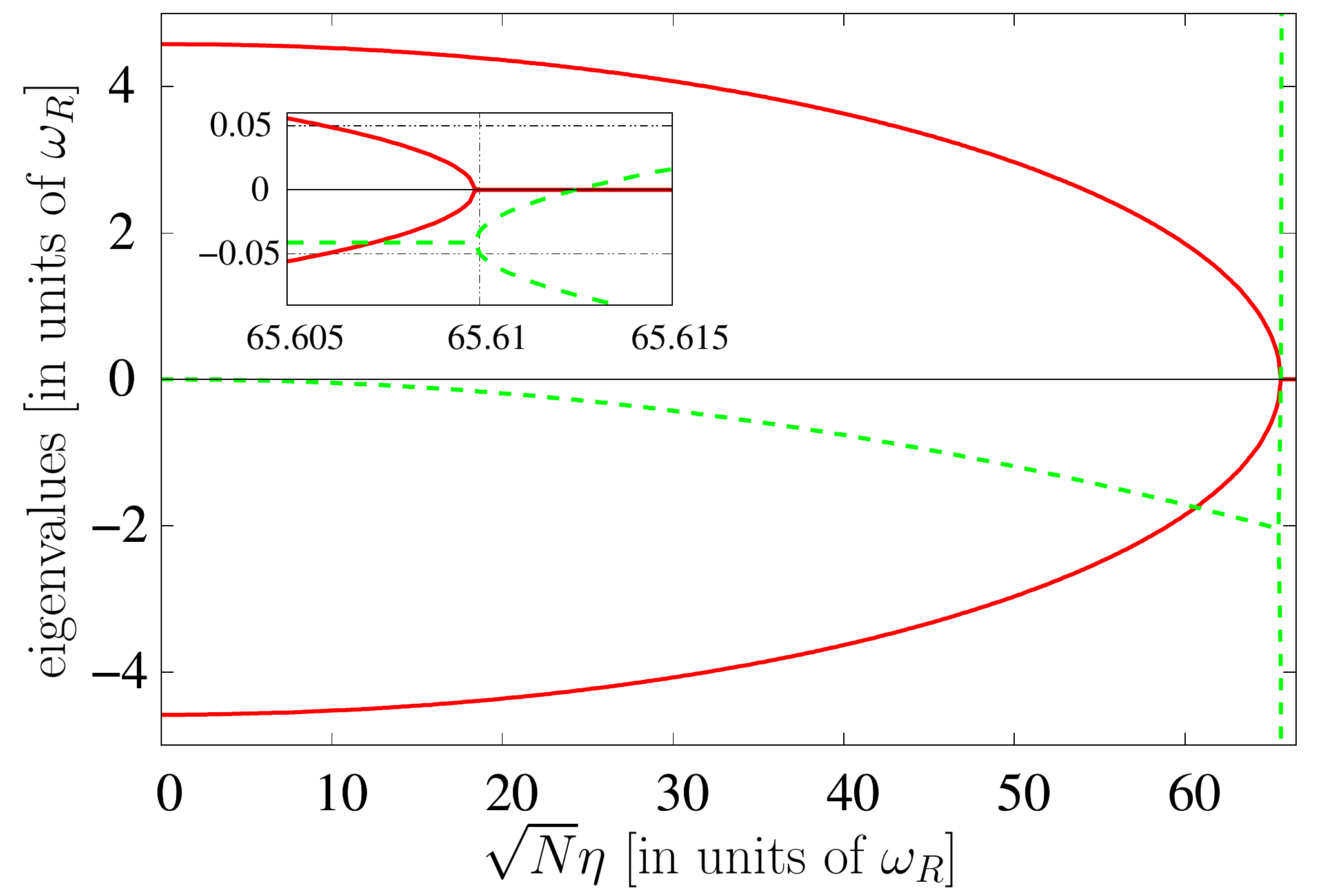}}}
\end{center}
\caption{Eigenvalues of the lowest condensate excitation of a homogeneous BEC from Eq.~(\ref{eq:kar}) as a
  function of the pump strength $\sqrt{N}\eta$. As shown in the inset, the
  real parts (solid red) vanishes slightly below the critical point which
  is reached when the upper branch of the imaginary parts (dashed
  green) crosses zero. In the main figure the imaginary parts are
  magnified by a factor of $50$. Parameters are the same as in
  Fig.~\ref{fig:theta_eta}.}
\label{fig:inst}       
\end{figure}
In Fig.~\ref{fig:inst}, we plot the numerical solutions of Eq.~(\ref{eq:kar}) for the lowest, dominantly
condensate-type excitation. The real parts (solid red) tend to zero as
increasing the pump strength. Oppositely, the absolute value of the imaginary parts of the eigenvalues (dashed green) increases with the pump strength. The behaviour of the eigenvalues near the critical point $\sqrt{N}\eta_c = 65.612\,\omega_R$ is magnified in the inset of  Fig.~\ref{fig:inst}. When the real parts reach zero at $\sqrt{N}\eta_*$,  the initially identical imaginary parts split up and the upper branch crosses zero. This crossing is the critical point, here the $\psi_0$ steady-state becomes  \emph{dynamically unstable}. By expressing $\eta$ from Eq.~(\ref{eq:kar}) for $\lambda = 0$, the same critical transverse pump amplitude $\eta_c$ is obtained as that in Eq.~(\ref{eq:eta_c}).

The emergence of a positive and a negative imaginary part just
at the vanishing of the real parts of an excitation is a typical
feature of a condensate's instability. In our case, however, coupling 
to the cavity field yields a negative imaginary part where the real part becomes zero. Thus, there is a narrow range, as shown by the inset of Fig.~\ref{fig:inst}, where the real parts remain zero, but both
imaginary parts are negative. This unfamiliar course of the dynamical instability is a signature of cavity cooling, thus it is so for finite cavity decay rate $\kappa$. 

In the limit $\Omega_1^2 \ll  \kappa^2 +\delta_C^2$, the first excitation frequency can be approximated by 
\bsubeq
\label{eq:anexc}
\be
\label{eq:exc1re}
\re\lambda_1 =  \Omega_1\sqrt{1 - \frac{\eta^2}{\eta_c^2}}\,, 
\ee
and the imaginary part is quadratic in the pump strength,
\be
\im\lambda_1 = -\frac{\kappa\,\Omega_1^2}{\delta_C^2+\kappa^2}\,
\frac{\eta^2}{\eta_c^2}\,.
\ee 
\esubeq
These approximate expressions fit well the curve shown in Fig.~\ref{fig:inst}. The s-wave interaction between atoms increases the decay rate of this particular excitation. The imaginary part vanishes, of course, for $\kappa \rightarrow 0$. 

The width of the narrow range, where the $\psi_0=1$ uniform steady state is stabilized by that the cavity cooling damps out zero-energy excitations, is obtained:
\be
\label{eq:eta_diff}
\eta_c^2 - \eta_{*}^2 =
\eta_c^2\left(\frac{\kappa\,\Omega_1}{\delta_C^2+\kappa^2}\right)^2. 
\ee
Similarly to the decay rate $\im\lambda_1$, 
the range expands on increasing the collisional parameter $Ng_c$. Finally, note that Eq.~(\ref{eq:eta_diff}) is considered  a small parameter in the estimation of $\lambda_1$ in  Eqs.~(\ref{eq:anexc}).

\subsection{Full spectrum}

Above threshold, the collective condensate-cavity excitations belong to the self-consistent ground state given by Eqs.~(\ref{eq:statsol}) which can be calculated only numerically.  Therefore, the solution of the eigenvalue
problem in Eq.~(\ref{eq:eigenproblem}) is performed also numerically using LAPACK. The wave function is defined on a spatial grid of 200 points  in the interval of one wavelength.

\begin{figure}
\begin{center}
\resizebox{0.90\columnwidth}{!}{%
\rotatebox{0}{\includegraphics{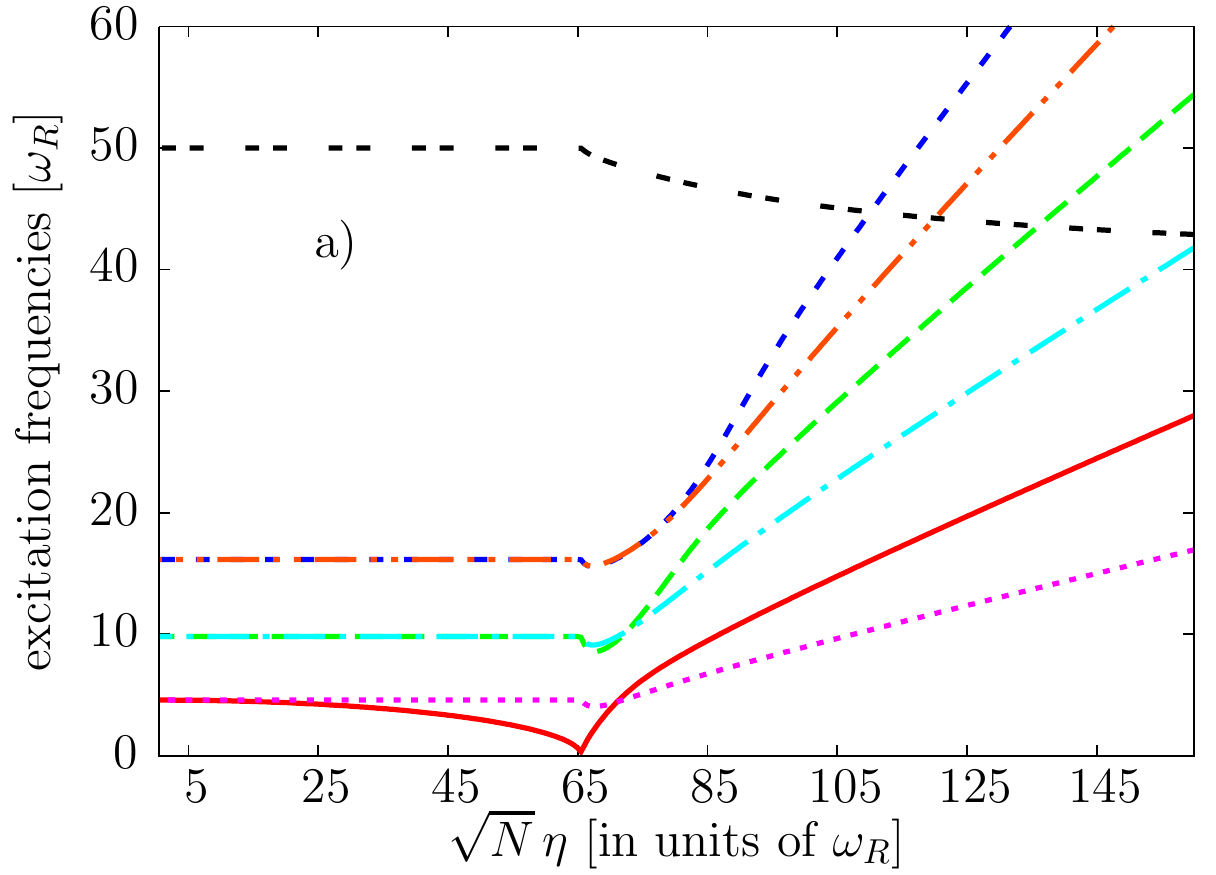}}}
\resizebox{0.90\columnwidth}{!}{%
\rotatebox{0}{\includegraphics{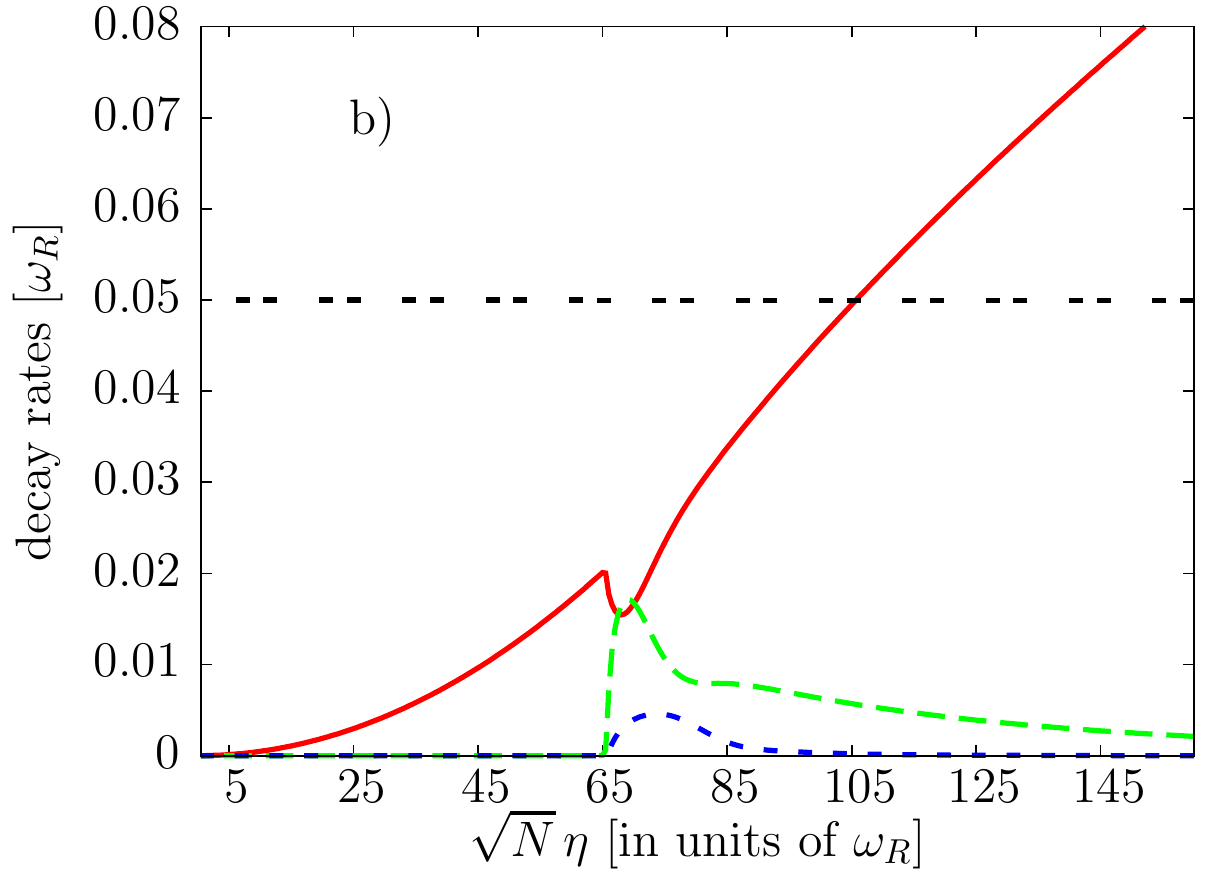}}}
\vspace{.3cm}
\end{center}
\caption{Frequencies (a) and decay rates (b) of the six lowest
  collective condensate excitations and the ones corresponding to 
  the cavity field as a function of the transverse pump strength.
  Decay rates of the lower branches of each
  pair are $0$.  For the cavity-dominated mode the frequency and the decay rate is divided by 5 and 4000, respectively, and $\gamma_1$ is divided by 2 in (b).  Parameters are the same as in Fig.~\ref{fig:theta_eta}.}
\label{fig:spectrum}       
\end{figure}

In Fig.~\ref{fig:spectrum}(a), we show the first three
excitation frequencies and damping rates ($\gamma_1/2$ is plotted) of
the condensate around the real ground state, and the frequency and damping rate of a higher excited state corresponding to the cavity mode (double
dashed lines, $\nu_f/5$ and $\gamma_f/4000$ are plotted). We depict
only the positive frequencies, taking into account the symmetry of the
eigenvalues of the matrix (\ref{mx:M}).

Below threshold the wave function $\psi_0(x)$ is constant, only the first excitation $\cos\,kx$ couples to the cavity field, and tends to zero on increasing the pumping, as it has been discussed before. All the other excitations are independent of the field, and have constant energies giving back
the Bogoliubov excitation spectrum of a condensate in a box, see Eq.~(\ref{eq:bogoliubov}). Two orthogonal excitations have a fixed number of nodes and are degenerate in this regime. One of the modes in each pair is orthogonal to the cavity mode function $\cos{kx}$,

At the onset of self-organization the eigenvalues dramatically change. At the critical point the excitation energies drop slightly below the  Bogoliubov energies given in Eq.~(\ref{eq:bogoliubov}). This dip is related to the collisions and disappears for $g_c\rightarrow 0$. Let us mention that the numerical calculation of the stationary solution of the Gross-Pitaeskii equation becomes inaccurate at the critical point. The convergence of the iterative solution slows down, which is an inherent consequence of the criticality obtained at the degeneracy of the ground and first excited state energies. Nevertheless we checked in some points that the dip is still there if very high accuracy is demanded in the iteration process.

At threshold, the degeneracy in the excitation pairs is lifted. The lower branch remains orthogonal to the cavity mode and decouples from the field fluctuations. The upper branches correspond to polariton excitations mixing condensate and field fluctuations. 

Far above threshold,  $\sqrt{N}\eta\rightarrow \infty$, the excitation frequencies increase linearly with $\sqrt{N}\eta$ and are uniformly spaced. This can be understood if we approximate the deep optical trap by  a harmonic potential in the vicinity of the antinodes. The adiabatic potential in Eq.~(\ref{eq:adpot}), transformed by using $\cos{kx} \approx 1-(kx)^2/2$, gets a characteristic harmonic frequency proportional to the square root of the intracavity intensity.  Thus, from Eq.~(\ref{eq:I0}), the trap frequency is linearly proportional to $\sqrt{N}\eta$. The spectrum is composed of the integer multiples of this  frequency. This is obvious in the collisionless case, $g_c=0$, where the excitation frequencies are the same as for a single particle in a harmonic potential. The other extreme case, the Thomas-Fermi regime which we are closer to with the parameters of Fig.~\ref{fig:spectrum}, can also be treated. Here, for a three-dimensional harmonic trap the excitation frequencies are given by a linear combination  with integer coefficients of the three vibrational frequencies \cite{stringari96}. This can be contracted to one dimensional motion by assuming very large frequencies in the transverse directions, then the low excitations are obtained as integer multiples of the longitudinal trap frequency.

In Fig.~\ref{fig:spectrum}b, the decay rate of the excitations corresponding to the upper branches in Fig.~\ref{fig:spectrum}(a) are shown. The lower branch of each pair has zero damping because it is
orthogonal to the mode function of the cavity field. The lowest excitation decays with a rate
$\gamma_1$ which exhibits a dip at threshold. In principle, it should drop down to zero, as shown in Fig.~\ref{fig:inst} representing the exact result below threshold. The present result, based on the numerical calculation of the ground state, is not accurate enough in the vicinity of the threshold to resolve the vanishing of $\gamma_1$. However, somewhat above threshold the numerical approach becomes very accurate. It shows then that $\gamma_1$ rises back to the value it had below threshold and then increases further with increasing $\sqrt{N}\eta$.  Not too far above threshold, the weakly localized ground state enables the coupling of higher condensate excitations to the cavity field. These modes become damped, but their damping rate vanishes in the $\sqrt{N}\eta\rightarrow \infty$ limit. There, as the ground state tends to a tightly localized one in a harmonic trap, only the second excitation  is coupled to the field \cite{horak01b}. 

The frequency and the decay rate of the cavity field are weakly
perturbed by the condensate. The frequency of the cavity mode is
expected to be $\nu_f = \Delta_C - NU_0 {\mathcal B}$,
that depends on the collective coupling $NU_0$ and, through the bunching parameter,  on the localization
of the ground state $\psi_0$. Decrease in the field mode frequency
is accompanied by an increase of the stationary photon number $I_0$ in the
cavity.

\section{Strong collective coupling regime}
\label{sec:defects}

The interpretation of self-organization in terms of adiabatic optical potentials in Sec.~\ref{subsec:selforg} reveals that the effect relies on the discrimination between the sites $k x=0$ and $k x=\pi$ (even and odd sites). The symmetry breaking is attributed entirely to the  $\lambda$ periodic potential. The other, $\lambda/2$ periodic $\cos^2kx$ potential could have been disregarded in this respect. Above threshold, however, this latter potential can have a significant role. 
If the condensate localizes around, say, $k x=0$, the $\cos\,kx$ potential has maxima at $k x=\pi$ and atoms are repulsed from this region, which assists the self-ordering process. The  $\cos^2kx$ potential term has always minima both at $k x=0$ and $k x=\pi$, thus it counteracts the repulsion. For certain parameters it may occur that a secondary
minimum appears and the adiabatic potentials form an asymmetric double well potential.  So called `defect' atoms scattering photons with `wrong' phase can be trapped in the shallower traps.

\begin{figure}[ht]
\begin{center}
\resizebox{0.95\columnwidth}{!}{%
\rotatebox{0}{\includegraphics{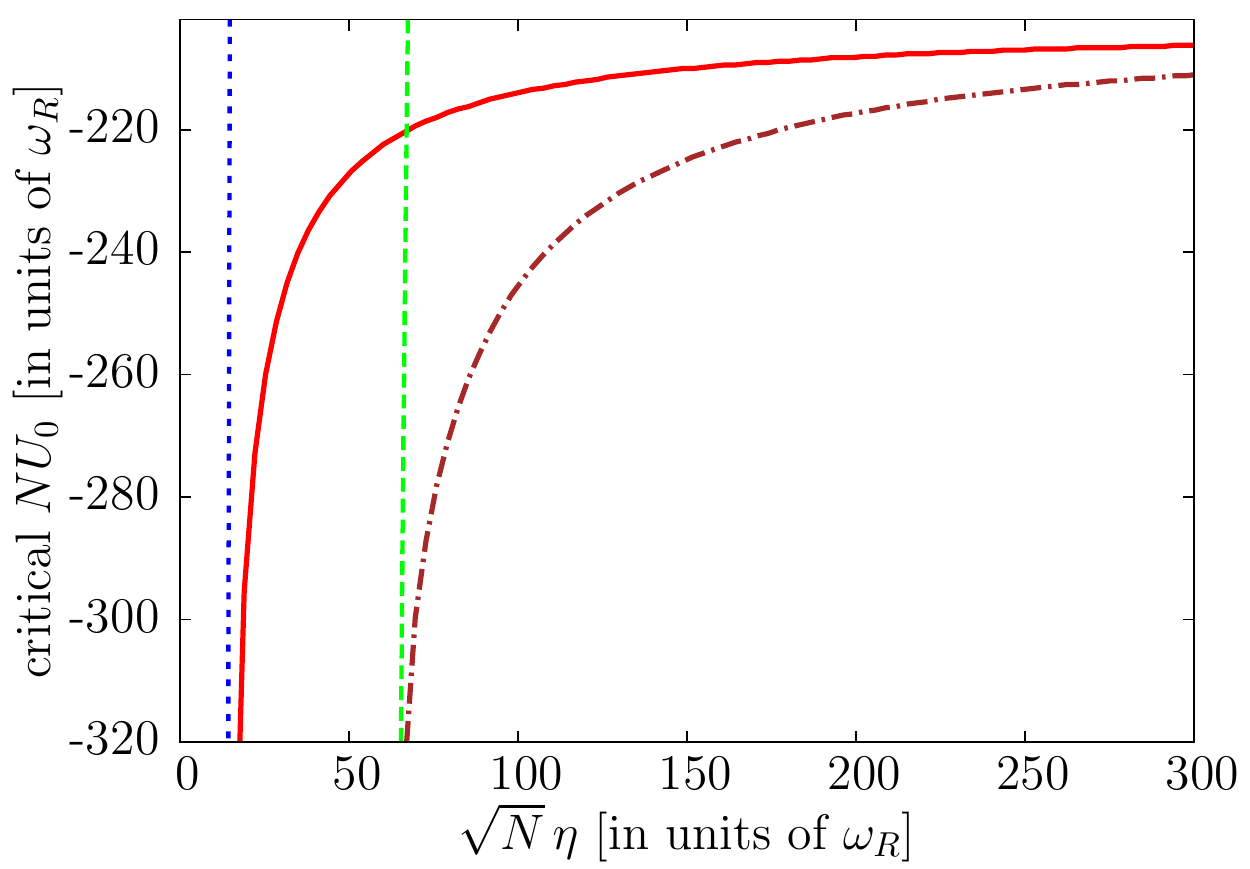}}}
\end{center}
\caption{Full phase diagram of the system for collision parameters $Ng_c = 0$ and $10$ $\lambda\omega_R$. For $Ng_c = 0$ the dashed blue curve represents the phase boundary between homogeneous distribution and self-organized lattice. In the region between the dashed blue and the solid red line (these curves coalesce asymptotically for large $NU_0$) there is a self-organized lattice without defects. The solid red curve gives the critical $N U_0$ below which the secondary minima occur.  Right shifted, the same phase boundary lines are drawn by dotted green and dashed-dotted brown for $Ng_c=10\lambda\omega_R$. Parameters are
  $\kappa = 200\omega_R$, $\Delta_C = - 2\kappa$.}
\label{fig:def_eta_g}       
\end{figure}

In the organized pattern, the emergence of a secondary potential minima at the complementary
sites depends on the ratio of the two potential terms in
Eq.~(\ref{eq:adpot}), namely on $U_1/U_2$. For perfect localization
($\eta \rightarrow \infty$), 
the condition for the possibility of stable defects is
\be
\label{eq:defcrit}
N|U_0| > \kappa \;,
\ee 
which is precisely the condition of strong collective coupling in cavity QED.
For finite pumping strengths ($N\eta < \infty$), the ratio of the two
potential terms is dependent on the order parameter $\Theta$, thus on
the localization of the condensate wave function. Therefore, the
above condition Eq.~(\ref{eq:defcrit}) should change with the pump
strength $N \eta$ and with the s-wave collision parameter $g_c$.
Because $U_2/U_1 \propto \Theta \leq 1$, the larger the order parameter 
$\Theta$ is, the smaller $N|U_0|$ is needed for getting defects. The critical lines are plotted in Fig.~\ref{fig:def_eta_g} below which
the secondary potential minima appear as a function of the pumping
strength for two different values of the s-wave collision parameter
$g_c=0$ and $N g_c=10 \lambda \omega_R$. The cavity detuning is fixed to $\Delta_C = -2\kappa$.  When collisions are not negligible, one needs a larger  light shift $N |U_0|$ to
reach the regime where defects are expected, just because the
collisions reduce the order parameter ${\Theta}$. The vertical asymptote of the
curves is at the threshold of the self-organization,
while the horizontal one is  at the same value defined by Eq.~(\ref{eq:defcrit}) that describes the $\sqrt{N}\eta
\rightarrow \infty$ behaviour for perfect localization.

\begin{figure}[ht]
\begin{center}
\resizebox{0.95\columnwidth}{!}{%
\rotatebox{0}{\includegraphics{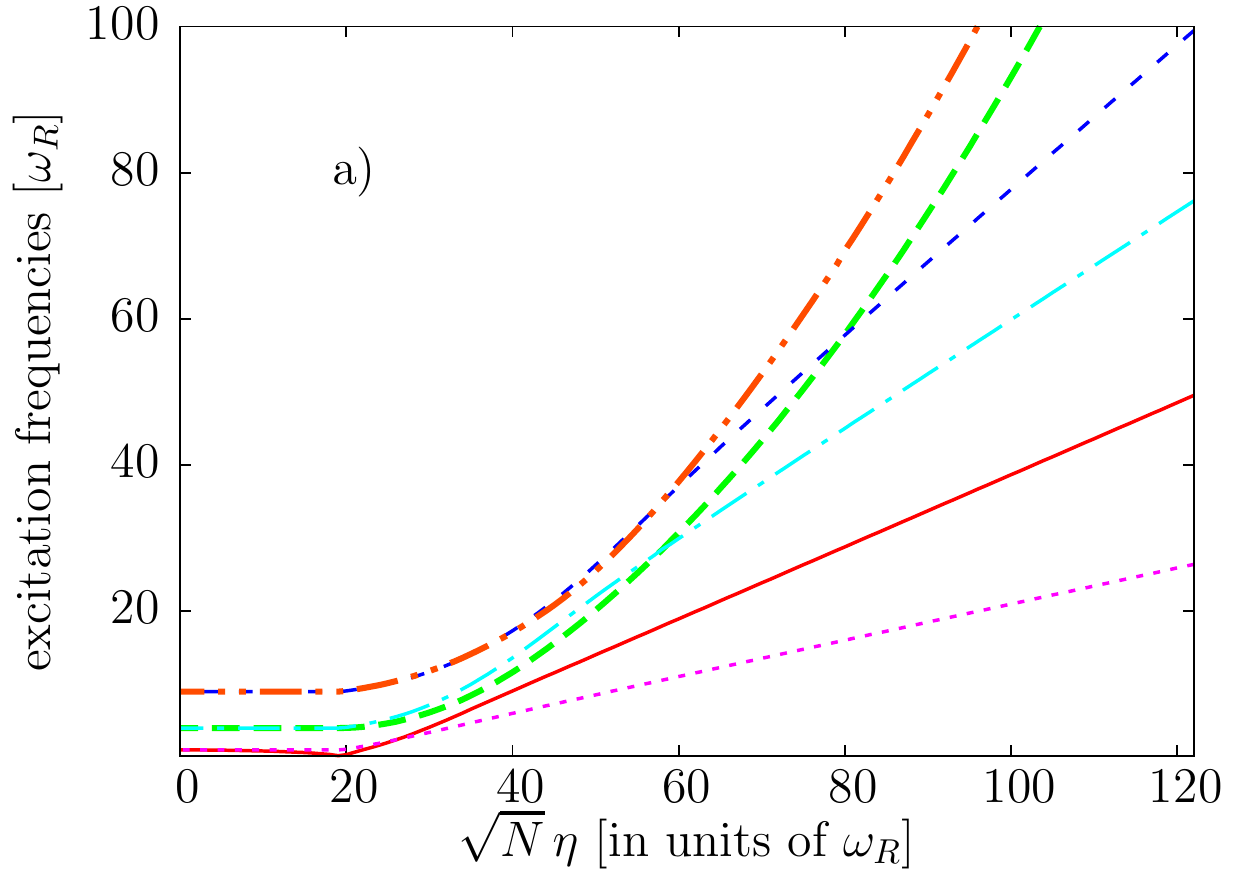}}}
\resizebox{0.95\columnwidth}{!}{%
\rotatebox{0}{\includegraphics{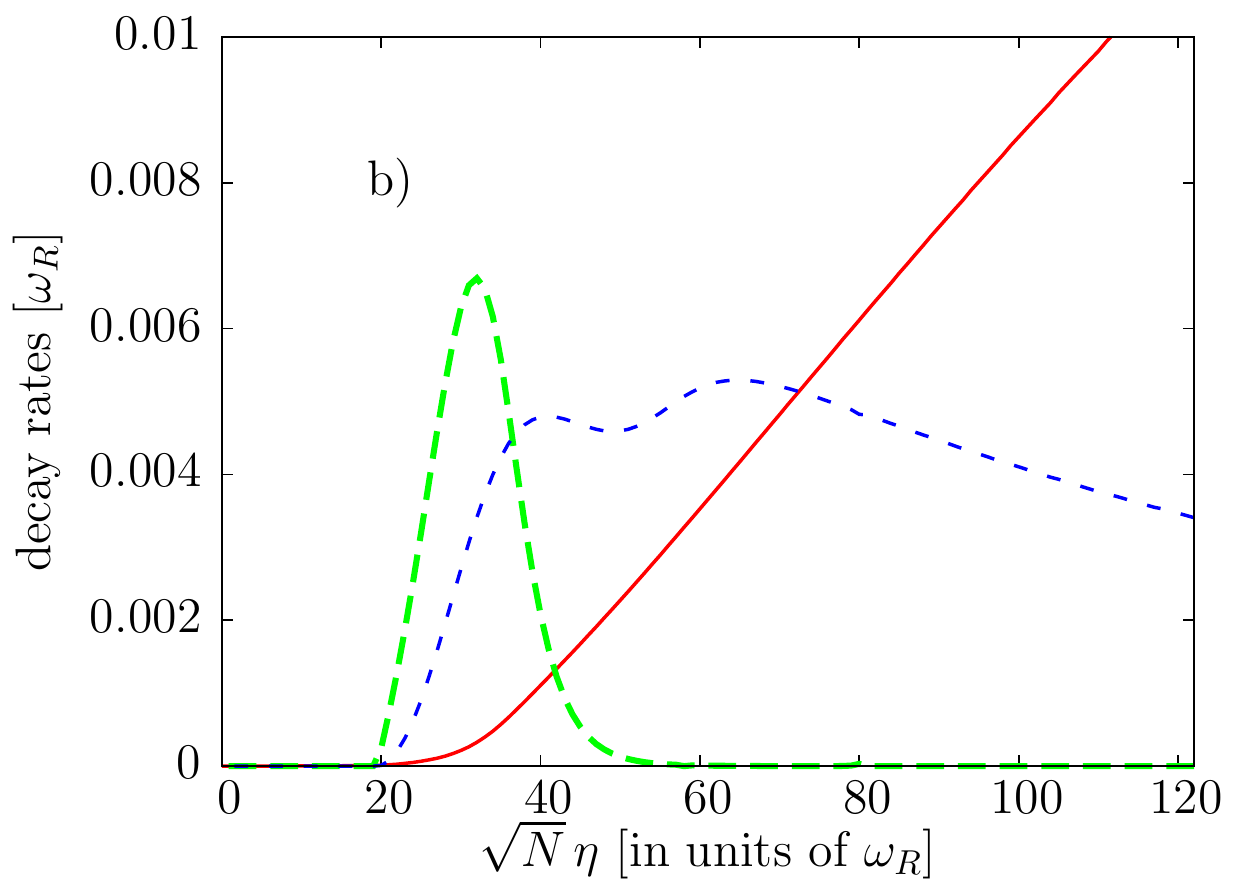}}}
\end{center}
\caption{Frequencies (a) and decay rates (b) of the six lowest
  collective condensate excitations in the regime, where the secondary
  potential minima are present as a function of the transverse pump
  strength. Decay rate of the first excitation $\gamma_1$ (solid red) is divided
  by $40$. Parameters are $g_c = 0$, $NU_0 = -1000\,\omega_R$, $\kappa =
  200\,\omega_R$ and $\Delta_C = -1200\,\omega_R$.}
\label{fig:spectr_U0_1}       
\end{figure}

Now let us turn to the effect of the secondary potential minima on the collective 
excitation spectrum. In the regime where the energy difference between the two types
of potential minima at the complementary sites is in the order of their depths,
the lowest excitation energies in the deeper wells are comparable to those
in the secondary ones. There will be collective excitations that are composed of
the combination of excitations localized in the two different wells. 

For simplicity, we calculate the spectrum in the collisionless case for $g_c = 0$.
In Fig.~\ref{fig:spectr_U0_1}, we plot the collective excitation frequencies (a) and 
the decay rates (b) as a function of the pumping strength $\eta$ for
$\kappa = 200\omega_R$ and $NU_0 = -1000\omega_R$. Starting from the Bogoliubov
spectrum of the homogeneous condensate, there appear two types of excitation
modes in the self-organized phase. The excitations that are localized in the deeper
wells are the same as the ones presented in Fig.~\ref{fig:spectrum}. Notice the lack of the dip at the critical point, which is because of the choice $g_c=0$.
The other type, represented by the two excitation modes bending up in Fig.~\ref{fig:spectr_U0_1}a contain 
the excitations of the `defect' well. For perfect localization, both  
the lower and the higher potential minima, hence also their difference, are proportional to $N|\eta|^2$. Exciting the condensate into the defect well costs the energy difference, this is the reason of the quadratic behavior as a function of the pumping strength $\sqrt{N}\eta$.  

\begin{figure}
\begin{center}
\resizebox{0.95\columnwidth}{!}{%
\rotatebox{0}{\includegraphics{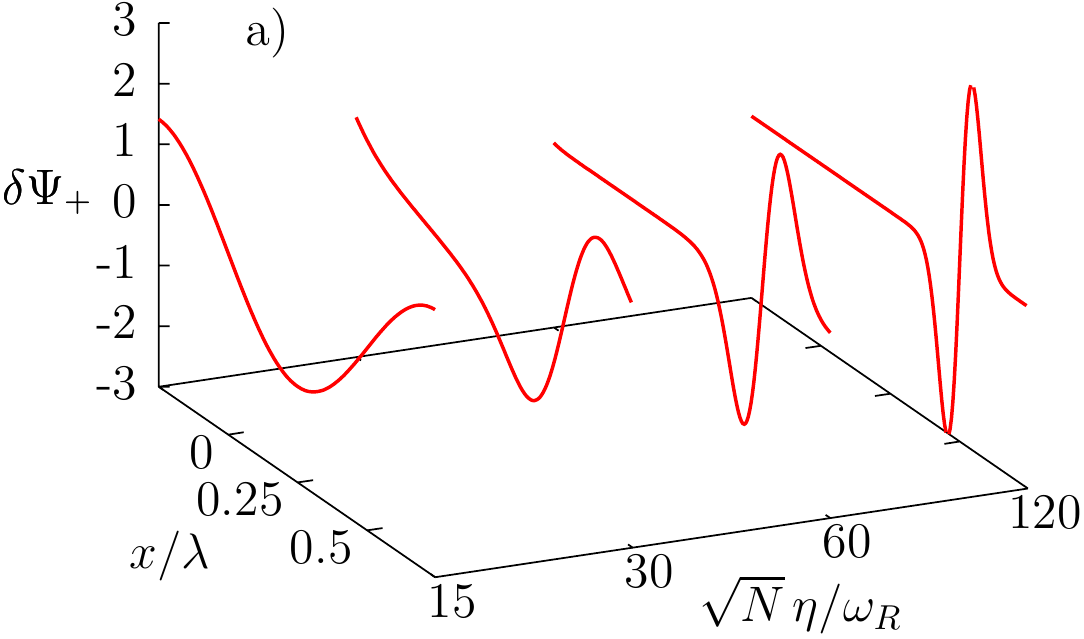}}}
\resizebox{0.95\columnwidth}{!}{%
\rotatebox{0}{\includegraphics{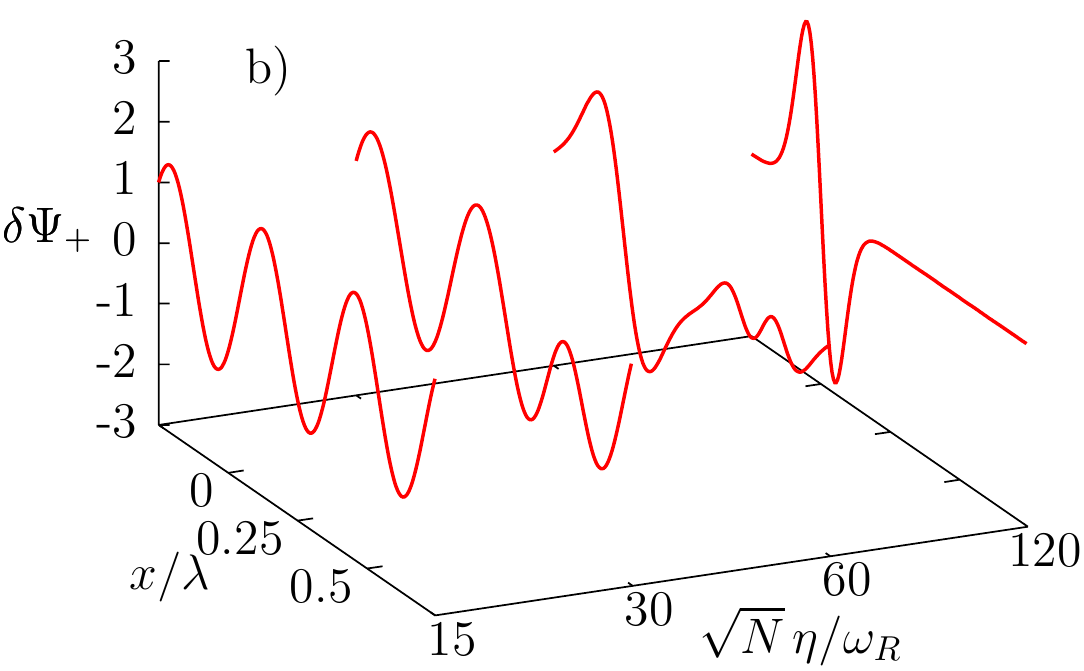}}}
\end{center}
\caption{The $\delta\psi_+(x)$ components of the eigenvectors of the collective
         excitations with the lowest (a) and the highest frequencies (b) 
         in Fig.~\ref{fig:spectr_U0_1}a are plotted for chosen values of 
         $\eta$. In these special cases the other components are zero.}
\label{fig:eigenvecs}       
\end{figure}

In order to exemplify the effect of the secondary potential minima in the spectrum,  we plot the wavefunctions of two condensate excitations at increasing values of the pump
strength $\sqrt{N}\eta$ in Fig.~\ref{fig:eigenvecs}. Namely, we chose the excitations with the lowest and
the highest frequencies in the right of Fig.~\ref{fig:spectr_U0_1}a. Both of them are decoupled from
the cavity field so that the wavefunction is real and can be interpreted as a condensate wavefunction in position space.  
The lowest one (a) is proportional to $\sin\,kx$ in the uniform phase
at $\sqrt{N}\eta = 15$, having two nodes at $x = 0$ and $\lambda/2$. On
increasing the pump strength $\sqrt{N}\eta$, the condensate wave function gets
more and more localized, hence this excitation contracts into the
primary potential well, which is now centered at $x = \lambda/2$ in
the self-organized phase.  At $\sqrt{N}\eta = 120$, the strongly localized BEC
feels just the harmonic term of the optical potential. Therefore, at
this end, we get for the wavefunction of the excitation something close to the first excitation of a harmonic oscillator. The other selected excitation is the highest one, bending
up in Fig.~\ref{fig:spectr_U0_1}a.  It has six nodes in the
homogeneous phase, that corresponds to the third excitation of the
Bogoliubov spectrum. It runs side by side with its orthogonal pair up 
to $\sqrt{N}\eta \approx 60$, where they split up. The lower branch tends to
the fourth harmonic oscillator excitation in the primary well for
$\sqrt{N}\eta \rightarrow \infty$, however the upper branch plotted in
Fig.~\ref{fig:eigenvecs}b becomes the first excitation in the
secondary well centered at $x = 0$.

\section{Quantum depletion}
\label{sec:depletion}

A crucial issue on the steady-state solution of the coupled
Gross--Pitevskii equations (\ref{eq:motion}) (besides the linear
stability analysis of the previous sections) is the magnitude of the
quantum depletion which is a small parameter in our model theory. In
our situation depletion is caused by two physical processes. 1) The
atom-atom interaction kicks out atoms from the condensate, such as in
a conventional interacting Bose gas. 2) The atom-light interaction
manifesting itself in the polariton like resonances prevents the
simple product wavefunction of the Gross--Pitaevskii theory to be a
true eigenstate of the coupled system. To synthesize the quantum
depletion caused by the polariton nature of some of the excited states,
we perform its calculation with the interatomic collision set to zero
($g_c=0$). We restrict this study to the case of a lossless cavity ($\kappa=0$), so that the
system be conservative and the number of noncondensed atoms ($N'$) can
be calculated in the usual way, based on the eigenvectors of the
Bogoliubov matrix $\mathbf{M}$ \cite{castin}.

\begin{figure}[ht]
\includegraphics[width=0.95\columnwidth]{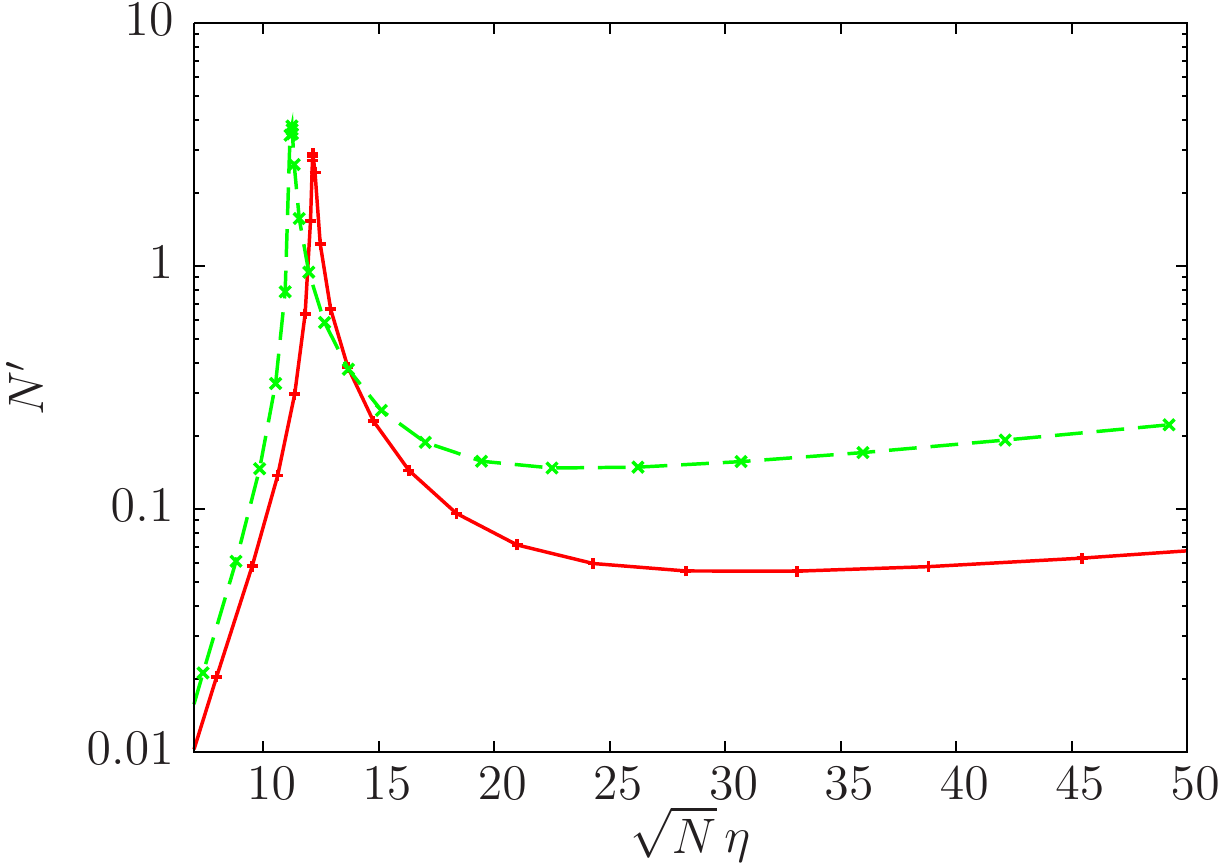}
\caption{Quantum depletion due to the coupling to the cavity field as a function of the pumping strength in a semi-logarithmic plot. $N^\prime$ is the number of atoms distributed in the excited states, for fractional depletion the plotted number has to be divided by the actual number of atoms $N$. Collisions are discarded, $g_c=0$. Other parameters: $\kappa=0$, $\Delta_C=-300$, $N U_0=-10$ (solid red curve) and $N U_0=-100$ (dashed green).}
\label{fig:depletion}
\end{figure}

Without the detailed calculation, we present the result for the quantum depletion in Fig.~\ref{fig:depletion}.
The sharp peak in the quantum depletion indicates the critical point. It is not surprising that the depletion increases up at the critical point where the first excited state becomes degenerate with the ground state. In fact, the number of atoms in the first excited state diverges as the pumping $\sqrt{N} \eta$ tends to the critical point from below as
\be
 N^\prime= \frac{\omega_R}{8 \lambda_1}
= \frac{1}{8} \left(\frac{\omega_R \delta_C}{8 N (\eta_c- \eta)^2}\right)^{\frac{1}{4}} \; .
\ee
where we used  Eq.~(\ref{eq:exc1re}) to approximate the first eigenvalue $\lambda_1$ near the threshold. 
The mean field theory and the perturbation approach is valid if the depletion rate $N^\prime/N$ is much smaller than 1. With reasonable experimental values of $N$, {\it i.e.} $N\gg 1$, this condition is obeyed with the exception of a very narrow range around the critical point. This justifies the use of the mean field model in almost the full range of Fig.~\ref{fig:depletion}. Moreover, by means of the divergence, the present model is suitable to locate precisely the critical point. More details on the quantum depletion and its treatment in an open system with $\kappa\neq 0$ will be the subject of a forthcoming paper.

\section{Conclusion}
\label{sec:conclusion}

Quantum many-body systems formed by neutral atoms weakly interacting through the electromagnetic field is of general interest. One possibility consists in using a dipolar gas \cite{stuhler:150406}. Although, the dipole-dipole interaction is relatively weak and of short range, patterns, collective modes \cite{ronen:013623} and instabilities \cite{ronen:030406}  arise in a dipolar condensate. The other possibility is that the atoms with tiny induced dipole moment are placed inside a cavity and couple strongly to one given mode of the resonator. Then the interaction has a long range, all atoms are coupled to all the other within the cavity. 
In this system, for example, the realization of the Dicke-model quantum phase transition is proposed and seems feasible \cite{dimer07,chen07,zhang07} by means of using Raman-type transitions in multilevel atoms. Another example, which concerns the motional degrees of freedom of particles, is that a thermal gas of laser-driven atoms produces a classical phase transition called `self-organization' in a cavity \cite{domokos02b}. In this paper we addressed the quantum version of this latter problem and gave a detailed account of the steady-state and the dynamics  when the atoms form a Bose-Einstein condensate in the cavity.

Within the framework of a Gross-Pitaevskii like mean-field model, we showed that the steady-state of the driven condensate is either the homogeneous distribution or a $\lambda$-periodic ordered pattern, and the two regimes are well separated by a critical point. That is, the quantum analogue of the classical self-organization phase transition exists for Bose-condensed atoms. The critical point, corresponding to a threshold pump power has been determined analytically from the Gross-Pitaevskii equation, and also analytically from the spectrum of the excited states. We showed that the spectrum of the collective excitations becomes more intricate in the strong collective coupling regime of cavity QED, where the ordered phase is described by an asymmetric double well potential and defect atoms can appear. We showed that the quantum depletion diverges in the critical point.  Otherwise, outside a narrow range around it, the coupled condensate-cavity system is stable and the mean-field approach is well justified.  


We acknowledge funding from the National Scientific Fund of Hungary (NF68736, T046129, and T049234).

\bibliography{cqed}

\end{document}